        \documentclass[%
         reprint,
        nofootinbib,
         amsmath,amssymb,
         aps,
        ]{revtex4-2}
        
        \usepackage{graphicx}
        \usepackage{dcolumn}
        \usepackage{bm}
        
        \usepackage[T1]{fontenc}
        \usepackage{inconsolata}
        \usepackage{url}
        \usepackage{hyperref}
        \usepackage{color}
        \usepackage{graphicx}
        \usepackage[table,xcdraw]{xcolor}

        \usepackage{listings}
               
\begin{document}
        
        \preprint{APS/123-QED}
        
        \title{Dynamical Photon Spheres in Charged Black Holes and Naked Singularities}
        \author{Yaghoub Heydarzade}\altaffiliation[ ]{Department of Mathematics, Faculty of Sciences, Bilkent University, 06800 Ankara, Turkey}
         \email{yheydarzade@bilkent.edu.tr}
        \author{Vitalii Vertogradov}
         \altaffiliation[ ]{Physics Department, Herzen State Pedagogical University of Russia, 48 Moika Emb., Saint Petersburg 191186, Russia
        SPb branch of SAO RAS, 65 Pulkovskoe Rd, Saint Petersburg 196140, Russia}
         \email{vdvertogradov@gmail.com}

        \date{\today}
        
        \begin{abstract}
    To understand the nature of a black hole shadow in dynamical spacetimes, we construct an analytical model of a dynamical photon sphere in the context of the Bonnor-Vaidya spacetime. Comparing the resulting photon sphere radius with the one in Vaidya spacetime, we find that the charge always decreases the radius of the photon sphere. We also prove that a naked singularity in Bonnor-Vaidya spacetime, unlike the static Reissner-Nordstrom naked singularity, may cast a shadow, and as a result, it cannot be distinguished from a black hole through its shadow. 
        \end{abstract}

        \maketitle
        
        
        \section{Introduction}
        The spacetime around a black hole is strongly curved such that an unstable circular photon orbit can exist near its event horizon. This photon sphere forms a shadow that an observer can see in the sky. Recent observations revealed shadows of supermassive black holes in the center of our galaxy and $M87$ \cite{bib:et1, bib:et2, bib:et3, bib:et4}. The idea that this shadow could be observed was brought forward in the year 2000 in a pioneering paper by
        Falcke et al. \cite{bib:15}. Tsupko et all \cite{bib:ruller} showed that a shadow can be used as a cosmological ruler. A concern regarding using the shadow as a cosmological ruler has been discussed in the paper \cite{bib:vanozy2}. The black hole shadow can be used to distinguish different black hole models and even to find differences between general relativity and alternative theories of gravity~\cite{bib:vanozi}. A review of an analytical study of static black hole shadow can be found in 
         \cite{bib:tsupko_review}. Also, the influence of plasma on the observed size of a black hole has been studied in the paper \cite{bib:plasma}. The mentioned studies considered static black holes. However, the real astrophysical black holes are dynamic and they change their mass and charge during the accretion and emission processes. Thus, the real black hole should be described by a dynamical spacetime. A numerical method of calculating a dynamical black hole shadow has been developed in \cite{bib:understanding, bib:japan}, and there are only a couple of analytical models \cite{bib:tsupko_first, bib:germany}. The problem with the analytical model is that in general there is only one conserved quantity in the spherically symmetric black hole spacetime; the angular momentum-per-mass $L$. Thus, to find other conserved quantities, one needs to seek extra symmetries to reduce the second-order differential equations of motion to the first-order. Dynamical spacetimes do not possess the timelike Killing vector $\frac{\partial}{\partial t}$, and hence one should look for conformal Killing vectors. If one considers the equation
        \begin{equation}
        K_{i;j}+K_{i;j}=C(x^l)g_{ij},~~i,j,l=0,...,3
        \end{equation}
        then the vector $K^i$ is called the homothetic Killing vector if $C(x^l)\equiv const.\neq 0$, and conformal Killing vector if $C(x^l)$ is an arbitrary function of the coordinates $x^l$. If the spacetime admits the conformal Killing vector, then there is an extra conserved quantity along null geodesics. If the spacetime admits the homothetic Killing vector, then there is a conserved quantity not only along the null geodesics but also along the timelike ones, see the appendix. The Vaidya and Bonnor-Vaidya spacetimes admit the homothetic Killing vector. Thus, one can find an extra conserved quantity along null geodesics which can help to reduce the second-order differential equation of motion to the first-order one. 
        
       The shadow formation in Vaidya spacetime has been investigated in the paper \cite{bib:germany}. The Vaidya solution \cite{bib:vaidya} is one of the exact dynamical solutions of the Einstein field equations. This spacetime is considered in describing the exterior geometry of a radiating star \cite{bib:santos1985non}. It can be regarded as a dynamical generalization of the static Schwarzschild solution. This
        spacetime is widely used in many astrophysical applications with strong gravitational fields.  The horizon structure of this solution has been investigated in \cite{bib:nel_vaidya, bib:nel_surface}. The Vaidya spacetime can be extended to include both the null dust and null string fluids leading to
        the generalized Vaidya spacetime \cite{bib:vunk}. A detailed investigation of the properties of the generalized Vaidya spacetime can be found in \cite{bib:husain1996exact, bib:Radiationstring1998, bib:twofluidatm1999}. The surrounded Vaidya spacetimes with cosmological fields have been studied in \cite{bib:tur1, bib:tur2, bib:tur3}. Also, a generalized Vaidya spacetime in the context of K-essence has been considered in a series of papers \cite{bib:manna1, bib:manna2, bib:manna3, bib:manna4}.   The charged generalization of the Vaidya solution was introduced by Bonnor and Vaidya \cite{bib:bonor}, and it has been widely investigated in gravitational collapse and naked singularity formation \cite{bib:ver_structure, bib:charged_collapse1, bib:charged_collapse2, bib:charged_collapse3}. The conformal symmetry properties and Hawking radiation of the Bonnor-Vaidya spacetime have been studied in \cite{bib:kudr, bib:charged_conformal, bib:maharaj_conformal, bib:charged_radiation}.  Also, the energy extraction process from the Bonnor-Vaidya black hole has been considered in \cite{bib:ver_extraction}. 
        
        The gravitational collapse of a massive star can lead not only to black hole formation but also to naked singularity \cite{bib:joshi_book, bib:joshi_review, bib:dvedy}. The gravitational collapse can also lead to an {\it eternal} naked singularity formation\footnote{Under the notion {\it eternal}, we mean that the naked singularity might be formed during the gravitational collapse and will never be covered with an apparent horizon.}  \cite{bib:joshi_eternal, bib:ver_eternal, bib:ver_structure}. The gravitational collapse in Vaidya spacetime can lead to an {\it event-like} naked singularity. However, in Bonnor-Vaidya spacetime the result of this process might be an {\it object-like} naked singularity\footnote{Under the {\it event-like} naked singularity we mean that  it is naked only at the event $v=0, r=0$. The notion {\it object-like} naked singularity means that the naked singularity might exist not only at the time $v=v_0$ but it can exist for period of the time $\delta v$. See, for example, explanations in~\cite{bib:joshi_review}.}. The analytical model of shadow formation in the dynamical Bonnor-Vaidya black hole hasn't been considered so far. The static Reissner-Nordstrom spacetime when $Q^2>M^2$ contains a naked singularity, but it cannot cast a shadow. However, there are static spacetimes containing a naked singularity, and under some physically relevant conditions, they can cast a shadow \cite{bib:joshi_naked}. The analytical model of shadow formation in a dynamical spacetime containing the naked singularity also hasn't been investigated yet. If a naked singularity can cast a shadow during the gravitational collapse, then it cannot be distinguished from a black hole through its shadow. 
        
        In this paper, we obtain the location of the dynamical photon sphere for both the black hole and naked singularity in the Bonnor-Vaidya spacetime. We consider the linear mass and charge functions that endow the spacetime with a homothetic Killing vector. By transforming the metric to the conformally static coordinates, we obtain the radius $R_{ph}$ of the photon sphere and elaborate on the influence of charge on this radius compared with the Vaidya case. We also obtain the angular size of a shadow which can be seen by an observer in the region where the homothetic Killing vector is timelike. Thereafter, we prove that the naked singularity in Bonnor-Vaidya spacetime, under some conditions, can also cast a shadow.   
        The organization of this paper is as follows. In section II, we briefly review the black hole shadow models in both the Reissner-Nordstrom and Vaidya solutions. In section III, after transforming the Bonnor-Vaidya metric to conformally static coordinates, we discuss the influence of the charge on the radius of a photon sphere. In section IV,  we implicitly prove that a naked singularity in Bonnor-Vaidya spacetime can cast a shadow. We give our concluding remarks in section V.

        The system of units $c=G=1$, and  the metric signature $\{-\,, +\,, +\,,+\}$ is used throughout the paper.

        \section{Shadow in Reissner-Nordstrom and Vaidya spacetimes}
        
        To compare how the dynamical shadow differs from the one in the Reissner-Nordstrom static case, and how the dynamical charge influences this shadow, we briefly review black hole shadows in Reissner-Nordstrom~\cite{bib:rn1, bib:rn2, bib:rn3}  and Vaidya~\cite{bib:germany} spacetimes in the following subsections.
        
        \subsection{Shadow in Reissner-Nordstrom Solution}
        
        The Reissner-Nordstrom black hole of a mass $M$ and a charge $Q$ has the following form
        \begin{eqnarray}\label{eq:reissner}
        ds^2&=&-f(r)dt^2+f^{-1}(r)dr^2+r^2d\Omega^2, \nonumber \\
        f(r)&=&1-\frac{2M}{r}+\frac{Q^2}{r^2}.
        \end{eqnarray}
        Here, $d\Omega^2=d\theta^2+\sin^2\theta d\varphi^2$ is the solid angle line element on the unit two-sphere.
        This spacetime admits two Killing vectors $\frac{\partial}{\partial t}$ and $\frac{\partial}{\partial \varphi}$ leading to the conserved energy-per-mass $E$ and angular momentum-per-mass $L$ given by
        \begin{eqnarray} \label{eq:energy_reissner}
        E&=&f(r)\frac{dt}{d\lambda},\nonumber\\
        L&=&r^2\frac{d\varphi}{d\lambda},
        \end{eqnarray}
        Here, $\lambda$ is an affine parameter.
        The radial motion in the equatorial plane $\theta=\frac{\pi}{2}$ reads
        \begin{eqnarray} 
        \left(\frac{dr}{d\lambda}\right)^2+V_{eff}(r)=E^2, 
        \end{eqnarray}
        where
        \begin{eqnarray} \label{eq:radial_reissner}
        V_{eff}(r)=f(r)\frac{L^2}{r^2},
        \end{eqnarray}
        here $V_{eff}(r)$ is the effective potential. 
        
        The radius $r_{ph}$ of the photon sphere is defined by the following conditions
        \begin{eqnarray} \label{eq:condition_potential}
        V_{eff}(r_{ph})=E^2, \nonumber \\
        \frac{dV_{eff}}{dr}|_{r=r_{ph}}=0.
        \end{eqnarray}
        Applying the second condition to \eqref{eq:radial_reissner} gives
        \begin{eqnarray} \label{eq:photo_reissner}
         rf'(r)-2f(r)=0,  
        \end{eqnarray}
        which leads to
        \begin{eqnarray}
        r^{\pm}_{ph}=\frac{1}{2} \left( 3M\pm \sqrt{9M^2-8Q^2}\right).
        \end{eqnarray}
        We should omit the minus sign because $r_{ph}^-\leq r_+=M+\sqrt{M^2-Q^2}$ where $r_+$ is the outer event horizon. Substituting this radius $r_{ph}^+$ into the first condition from \eqref{eq:condition_potential}, one obtains
        \begin{eqnarray} \label{eq:impact_reissner}
        b_{rn}&=&\frac{r_{ph}^2}{\sqrt{Mr_{ph}-Q^2}}\nonumber \\
        &=&\frac{\left[1+\sqrt{1-\frac{8}{9}\sigma^2}\,\right]^2}
{2\sqrt{2}\sqrt{1+\sqrt{1-\frac{8}{9}\sigma^2}-\frac{2}{3}\sigma^2}}\,b_{\rm sch},
        \end{eqnarray}
        where $b^2\equiv \frac{L^2}{E^2}$ is defined as the impact parameter which regarding to the first Eq. in (\ref{eq:condition_potential}) is equal to $b^2\equiv \frac{L^2}{V_{eff}(r_{ph})}$ for the photon sphere radius. Here, $b_{rn}$ and $b_{sch}=3\sqrt{3}M$ are the impact parameters in Reissner-Nordstrom and Schwarzschild spacetimes, respectively,  and the dimensionless parameter $\sigma$ is defined as
        \begin{equation}
        \sigma \equiv \frac{Q}{M} \quad, \quad 0\leq |\sigma|\leq 1.
        \end{equation}
        Note, if and only if $Q=0=\sigma$ \footnote{We consider only the case of a black hole i.e. $M^2\geq Q^2$.} then $b_{rn}=b_{sch}$. From \eqref{eq:impact_reissner} one obtains the angular radius $\omega_{sh}$ of the shadow as can be seen an observer at radius $r_o$ from the black hole~\cite{bib:tsupko_review}
        \begin{equation} \label{eq:angular_reissner}
        \sin\omega_{sh}=\frac{\sqrt{r_o^2-2Mr_o+Q^2}}{r_o^2}b_{rn}.
        \end{equation}
        If the observer's position $r_o$ is far away from the black hole i.e. $r_o\gg r_+$ then the angular size of a black hole can be approximately given by
        \begin{equation}
        \omega_{sh}=\frac{b_{rn}}{r_o}\,.
        \end{equation}
        \subsection{Shadow in Vaidya spacetime}
        
        The in-going (advanced time) Vaidya solution ~\cite{bib:vaidya} in Eddington-Finklestein coordinate is given by the line element
        \begin{eqnarray} \label{eq:usualvaidya}
        ds^2&=&-f(v,r)dv^2+2dvdr+r^2d\Omega^2,\nonumber \\
        f(v,r)&=&1-\frac{2M(v)}{r},
        \end{eqnarray}
        where $v$ is Eddington’s advanced time coordinate, and $d\Omega^2=d\theta^2+\sin^2\theta d\varphi^2$ is the solid angle line element on the unit two-sphere. This spacetime admits the homothetic Killing vector for the mass function of the linear form~\cite{bib:nel_vaidya}
        \begin{equation} 
        M(v)=\mu v,
        \end{equation}
        where $\mu$ is a positive constant. One can introduce coordinate transformations  $(v, r)\to(t, R)$ by \cite{bib:germany}
        \begin{eqnarray}\label{eq:trans}
        v&=&r_0e^{\frac{t}{r_0}},\nonumber \\
        r&=&Re^{\frac{t}{r_0}},
        \end{eqnarray}
        where $t$ and $R$ are the new time and radial coordinates, respectively, and $r_0$ is an arbitrary constant of the length dimension. Then  one obtains the Vaidya spacetime in conformally-static coordinates as
        \begin{eqnarray} 
        ds^2&=&e^{\frac{2t}{r_0}}d\tilde{s}^2,
        \end{eqnarray}
        where
        \begin{eqnarray}\label{eq:vaidya_conformal}
        d\tilde{s}^2=-\left(1-\frac{2\mu r_0}{R}-\frac{2}{r_0}R\right)dt^2+2dRdt+R^2d\Omega^2.
        \end{eqnarray}
        The metric leading
to this line element represents a generalization of the Schwarzschild spacetime.
        One can find that the homothetic Killing vector $\frac{\partial}{\partial t}$ is timelike in the region of outer communication, i.e. in the region $R_-<R<R_+$ between inner $R_-$ and outer $R_+$ horizons, where
        \begin{eqnarray} \label{eq:timelikehkv}
        R_{\pm}=\frac{r_0}{4}\left(1\pm \sqrt{1-16\mu}\right).
        \end{eqnarray}
        The key point is that the homothetic Killing vector in the region $R_-<R<R_+$ introduces, in addition to conserved angular momentum-per-mass in equatorial plane $\theta=\frac{\pi}{2}$
        \begin{equation}
        L=R^2\frac{d\varphi}{d\lambda},
        \end{equation}
        one more constant of motion
        \begin{equation}
        E=e^{\frac{2t}{r_0}}\left[\left(1-\frac{2\mu r_0}{R}-2\frac{R}{r_0}\right) \frac{dt}{d\lambda}-\frac{dR}{d\lambda}\right].
        \end{equation}
        The radial component of the four-velocity then reads
        \begin{eqnarray} 
        \left(\frac{dR}{d\lambda}\right)^2+V_{eff}=E^2,
        \end{eqnarray}
        where
        \begin{eqnarray}\label{eq:radial_vaidya}
        V_{eff}=\left(1-\frac{2\mu r_0}{R}-2\frac{R}{r_0}\right)\frac{L^2}{R^2}.
        \end{eqnarray}
        Now, we can find the shadow that can be seen by an observer moving on $t$-lines in the region $R_-<R<R_+$ with \eqref{eq:timelikehkv}.
        To find the radius $R_{ph}$ of possible photon spheres, we use the second condition for the effective potential $V_{eff}$ in \eqref{eq:condition_potential}. This leads to two radii  $R^{\pm}_{ph}$ as
        \begin{equation} \label{eq:vaidya_condition1}
        R^{\pm}_{ph}=\frac{r_0}{2}\left(1\pm \sqrt{1-12\mu}\right)\,.
        \end{equation}
        Note that we do not consider the plus sign because it leads to the imaginary value of the impact parameter $b^2=\frac{L^2}{E^2}$. The minus sign leads to the Schwarzschild radius of a photon sphere in the limit of small $\mu$ as
        \begin{equation}
        R_{ph}^-=3r_0\mu+O(\mu^2) \,.
        \end{equation}
        Returning to the old coordinates $(r,v)$ one obtains
        \begin{equation}
        r_{ph}=3\mu v=3M(v),~~ \mu \ll 1.
        \end{equation}
        Now, substituting the radius $R_{ph}^-$ \eqref{eq:vaidya_condition1} into the first condition \eqref{eq:condition_potential}, one obtains
        \begin{equation} \label{eq:vaidya_condition2}
        b_V=\sqrt{\frac{R_{ph}^-}{4\mu r_0-R_{ph}^-}}R_{ph}^-,
        \end{equation}
        where $b_V$ is the impact parameter for the Vaidya spacetime. Note that if we substitute $R_{ph}^-\approx 3\mu r_0$ into \eqref{eq:vaidya_condition2} one obtains $b_V\approx 3\sqrt{3}\mu r_0$.
        The angular size of a shadow which can be seen by an observer at the radius $R=R_o$ is given by
        \begin{equation} \label{eq:vaidya_angular}
        \sin^2 \omega_{sh} =\frac{b_V^2\left(R_o-2\mu r_0 -\frac{2}{r_0}R_o\right)}{R_o^3}.
        \end{equation}
        Note that the right-hand-side of the equation \eqref{eq:vaidya_angular} tends to zero when the position of the observer tends to one of the conformal horizons, i.e. $R_o\rightarrow R_{\pm} \rightarrow \sin^2\omega_{sh}\rightarrow 0$. When the observer position coincides with the radius of a photon sphere $R_o=R^-_{ph}$ then the right-hand-side \eqref{eq:vaidya_angular} tends to unity $\sin^2 \omega_{sh}=1$. So, one can conclude that $0\leq \omega_{sh}\leq \frac{\pi}{2}$ when $R^-_{ph}\leq R_o\leq R_+$ and $\frac{\pi}{2}\leq \omega_{sh}\leq \pi$ when $R_-\leq R_0\leq R^-_{ph}$. We can consider the following cases
        \begin{itemize}
        \item The observer location near the horizon $R_o\rightarrow R_+$: In this case, $\omega_{sh}\rightarrow 0$, i.e. the observer sees a bright sky. In coordinates $(r,v)$ it corresponds to the observer position at
        \begin{equation} \label{eq:vaidya_f1}
        r_o=\frac{v}{4}\left(1+\sqrt{1-16\mu}\right)\approx \frac{v}{2}-2\mu v=\frac{v}{2}-r_{ah}(v)>0,
        \end{equation}
        where $r_{ah}=2\mu v$ is the apparent horizon location in Vaidya spacetime. 
        \item $R_o=R^-_{ph}$: In this case, the observer sees half of the sky $\omega_{sh}=\frac{\pi}{2}$. In old coordinates, it corresponds to the observer location at
        \begin{equation}
        r_o=\frac{v}{2}\left(1-\sqrt{1-12\mu}\right)\approx 3\mu v.
        \end{equation}
        \item $R_o\rightarrow R_-$: This case corresponds to a dark sky i.e. $\omega_{sh}\rightarrow \pi$. In this case, the observer, in coordinates $(r,v)$ is located at 
        \begin{equation} \label{eq:vaidya_f2}
        r_o=\frac{v}{4}\left(1-\sqrt{1-16\mu}\right) \approx 2\mu v=r_{ah},
        \end{equation}
        where $r_{ah}=2\mu v$ is the apparent horizon location in the Vaidya spacetime.  In this case, the observer is located near the apparent horizon of a black hole. The expression in \eqref{eq:vaidya_f2} corresponds to the quasi-local location of the event horizon in the Vaidya spacetime~\cite{bib:nel_vaidya, bib:kudr}.
        \end{itemize}
        \section{The influence of the charge on the black hole shadow}
        
        In this section, we will consider the influence of the charge on the dynamical photon sphere, and the findings here are all novel. We give some important findings as Remarks. 
        
        The generalization of the Vaidya spacetime in the presence of an electromagnetic source was introduced by Bonnor and Vaidya \cite{bib:bonor}. This solution has the following line element in the Eddington-Finkelstein coordinates $\{v, r, \theta, \varphi\}$
        \begin{equation} \label{eq:chargedvaidya}
        ds^2=-f(v,r)dv^2+2dvdr+r^2 d\Omega^2,
        \end{equation}
        where
        \begin{equation}
        f(v,r)=1-\frac{2M(v)}{r}+\frac{Q^2(v)}{r^2}.
        \end{equation}
        Here $v$ is Eddington's advanced time coordinate, $M(v)$ and $Q(v)$ are the black hole mass and electric charge, respectively, and  $d\Omega^2=d\theta^2+\sin^2\theta d\varphi^2$ is the solid angle line element on the unit two-sphere. The solution given in \eqref{eq:chargedvaidya} has the region $r < \dot Q Q/\dot M$ where the weak energy condition is violated \cite{bib:pois, bib:energy_condition}. However, this region is hidden with the apparent horizon and particles can't get there due to Lorentz force~\cite{bib:charged_no_violation}. 
        
        The black hole shadow problem arises when we consider the dynamical spacetime \eqref{eq:chargedvaidya} because, in general, we have only one constant of motion associated with the angular momentum per unit mass
        \begin{equation}
        L=r^2\frac{d\varphi}{d\lambda},
        \end{equation}
        which is not enough to reduce the second-order geodesic equations to the first-order.  For certain choice of the mass $M(v)$ and charge $Q(v)$ functions
        \begin{eqnarray} \label{eq:functions}
        M(v)&=&\mu v,~ \mu >0, \\
        Q^2(v)&=&\alpha \nu^2 v^2,~ 0<\alpha \ll 1.
        \end{eqnarray}
         the spacetime \eqref{eq:chargedvaidya} admits the homothetic Killing vector ~\cite{bib:maharaj_conformal,  bib:kudr, bib:charged_conformal}. Here $\mu$ and $\nu$ are positive constants, and $\alpha$ is a small dimensionless parameter that we later use to define the approximate conformal killing horizons and photon sphere. One should note that the above choice of mass and charge functions is the only possibility for having a homothetic Killing vector in Bonnor-Vaidya spacetime. This spacetime does not admit a homothetic Killing vector for other forms of these functions.
        Substituting these functions,  our line element takes the form
        \begin{eqnarray} \label{eq:metric}
        ds^2&=&-f(v,r)dv^2+2dvdr+r^2d\Omega^2, \nonumber \\
        f(v,r)&=&1-\frac{2\mu v}{r}+\frac{\alpha \nu^2v^2}{r^2}.
        \end{eqnarray}
         By doing the transformations (\ref{eq:trans}), the line element \eqref{eq:metric} takes the following form
        \begin{eqnarray} \label{eq:metconformal}
        ds^2=e^{\frac{2t}{r_0}}d\tilde{s}^2, 
        \end{eqnarray}
        where 
        \begin{eqnarray}
        \label{eq:bvmetric}
        d\tilde{s}^2&=&-\left(1-\frac{2\mu r_0}{R}+\frac{\alpha \nu^2 r_0^2}{R^2}-\frac{2}{r_0}R\right)dt^2\nonumber\\
        &&+2dtdR+R^2d\Omega^2.
        \end{eqnarray}
        Here $d\tilde{s}^2$ represents a generalization of the Reissner-Nordstrom spacetime which can be referred to as "specialized Kiselev spacetimes" \footnote{The metrics leading to the line elements (\ref{eq:vaidya_conformal}) and (\ref{eq:bvmetric}) are special cases of the Kiselev solution \cite{bib:50} where a linear term in the radial coordinate appears in the metric function. Some authors claimed such solutions as surrounded solutions by "quintessence". It was pointed out by Visser that this is not the case despite the over 150 papers having made similar claims \cite{mviss}. While the energy-momentum tensors for spacetimes given by (\ref{eq:metconformal}) and (\ref{eq:bvmetric}) are not isotropic and represent some combination of matter (imperfect fluids) and fields (scalar and electromagnetic) it is clear that it is not quintessence in the form employed in cosmological models with accelerated expansion. It is suggested that such metrics be referred to as "specialized Kiselev spacetimes".}.
We consider the following range of the coordinates $0\leq v <+\infty$ and $0\leq r <+\infty$ which corresponds to the following domains in new coordinates $-\infty<t<+\infty$ and $0\leq R<+\infty$.
        
        The homothetic Killing vector $K^i=v\frac{\partial }{\partial v}+r\frac{\partial }{\partial r}$ after the transformation \eqref{eq:trans} becomes
        \begin{equation}\label{eq:hkv}
        K^i = \frac{\partial}{\partial t}.
        \end{equation}
        This vector is not timelike everywhere, and we have the following remarks.  
        
         {\bf Remark 1:} If $\mu >\frac{1}{12}$, there is  the region  $0<R<R_c$ where $K^i$ is timelike. It means that the spacetime contains a naked singularity, not a black hole, and $R=R_c$ is the cosmological horizon \footnote{ Here $R_c$ is not the cosmological horizon in de Sitter space. It is called "cosmological" because of being an outer horizon possessing a large enough area in comparison to the inner horizon. Indeed, $R_c$ is the greatest real root of the equation $g_{00}=0$ with $g_{00}$ given in (\ref{eq:bvmetric}) depending on specific values of $\mu, \nu, \alpha$ and $r_0$. }.   Thus, we find the first constraint as $\mu\leq \frac{1}{12}$ if one is interested only in the black hole case. One should note that this constraint coincides with the one obtained in Vaidya spacetime \cite{bib:nel_vaidya, bib:germany}.
        
         {\bf Remark 2:} By choosing  parameters $\mu,\nu,~\alpha$ with the constraint $\mu \leq \frac{1}{12}$ to satisfy the equation \footnote{Here, we used the condition for a black hole to be an extremal one, i.e. $f(v, r_{h})=0$ and $f'(v, r_{h})=0$ in (\ref{eq:metric}) where the prime sign denotes the radial derivative.}
        \begin{equation} \label{eq:condition_extremal}
        2-36 \mu+33\sqrt{1-12\mu}+(1-12 \mu)^\frac{3}{2}+108 \alpha \nu^2 =0,
        \end{equation} 
        the homothetic Killing vector $K^i$ is timelike in the region $R\in \left(0,R_h\right) \cup \left(R_h, R_c\right)$ where $R_h=\frac{r_0}{6}\left(1-\sqrt{1-12 \mu}\right)$ is the conformal Killing horizon of the extremal black hole. So, we will consider this case for the observer located in the region $R_h<R<R_c$. From \eqref{eq:condition_extremal}, we obtain the second constraint $0 \leq \alpha \nu^2 \leq \frac{1}{108}$ because $\alpha >0$.
        
         {\bf Remark 3:} If the parameters $\mu, \nu, \alpha$ satisfy the inequality \footnote{We obtain these inequalities by solving the equation $f'(r_1)=0$ and then we demand $f(r_{1})<0$.}
        \begin{equation} \label{eq:condition_nonextremal}
        2-36 \mu+33\sqrt{1-12\mu}+(1-12 \mu)^\frac{3}{2}+108 \alpha \nu^2 <0,
        \end{equation} 
        and
        \begin{equation}\label{eq:condition_nonextremal2}
        2-36\mu-45\sqrt{1-12\mu}-(1-12\mu)^{\frac{3}{2}}+108 \alpha \nu^2 >0,
        \end{equation}
        the vector $K^i$ is timelike in the regions $R\in (0, R_-) \cup (R_+, R_c)$ where $R_{\mp}$ are the inner and outer horizons, respectively. In order to observe the photon sphere, the physical observer should be located at the region  $R\in \left(R_+, R_c\right)$. One notes that the condition \eqref{eq:condition_nonextremal2} is a key condition to have a black hole with two horizons (The conditions for an extremal black hole have been considered in the previous remark).
        
        {\bf Remark 4: } In Remark 2, we found the second constraint as $\alpha \nu^2\leq \frac{1}{108}$. However,  $\alpha \nu^2$ must be even smaller in the case of a black hole. For a black hole, the apparent horizon is given by
        \begin{equation}
        r_{ah}^+=\left(\mu+\sqrt{\mu^2-\alpha \nu^2}\right)v,
        \end{equation}
        and here one observes that to avoid the naked singularity formation, one must demand $\mu^2\geq \alpha \nu^2$. Hence, considering $\mu\leq \frac{1}{12}$, we find that the second constraint gets tighter as
        \begin{equation} \label{eq:second_constraint}
        \alpha \nu^2\leq \frac{1}{144}.
        \end{equation} The first observation is that if the conditions \eqref{eq:condition_nonextremal} and \eqref{eq:condition_nonextremal2} are satisfied, then we have three horizons $R_{\mp}, R_c$, where $R_{\mp}$ corresponds to the inner and outer horizons, respectively, and $R_c$  is interpreted as a cosmological horizon. The inner horizon $R_-$ is out of our interest and we will consider the observer located at $R=R_o$ between outer and cosmological horizons, i.e. $R_+\leq R_o\leq R_c$. If we compare these horizons with $R^V_{\mp}$ for the Vaidya solution, where $R_-$ is the black hole horizon and $R_+$ can be interpreted as a cosmological one,  then we find that
        \begin{equation}
        R_-^V>R_+^{BV},~~ R_+^V<R_c^{BV},
        \end{equation}
        where  $V$ and $BV$ stand for "Vaidya" and "Bonnor-Vaidya", respectively. \\
        {\bf Remark 5:} Here one concludes that the region where the homothetic Killing vector $\frac{\partial }{\partial t}$ is timelike, is bigger than in the uncharged Vaidya case. 
        
        The metric in conformally static coordinates admits two constants of motion
        \begin{eqnarray} \label{eq:charged_energy}
        E&=&e^{\frac{2t}{r_0}}\left[\left(1-\frac{2\mu r_0}{R}+\frac{\alpha \nu^2 r_0^2}{R^2}-2\frac{R}{r_0}\right)\frac{dt}{d\lambda}-\frac{dR}{d\lambda} \right],\nonumber \\
        L&=&e^{\frac{2t}{r_0}}R^2\frac{d\varphi}{d\lambda}.
        \end{eqnarray}
        Using these constants of motion, one can obtain the equation for radial motion in the form
        \begin{eqnarray} \label{eq:charged_potential}
        &&\left(\frac{dR}{d\lambda}\right)^2+V_{eff}=E^2,\nonumber \\
        &&V_{eff}=\left(1-\frac{2\mu r_0}{R}+\frac{\alpha \nu^2 r_0^2}{R^2}-2\frac{R}{r_0}\right)\frac{L^2}{R^2}.
        \end{eqnarray}
        We can follow the same method of defining a photon sphere described in the previous section. However, the second condition \eqref{eq:condition_potential} gives the cubic equation here and the solutions are too cumbersome. Instead, we note that from \eqref{eq:second_constraint} we have obtained that $\alpha \nu^2$ is small and, assuming $\alpha \ll 1$, we can consider the radius of a photon sphere as
        \begin{equation} \label{eq:charged_radius}
        R_{ph}=R_{ph}^{(0)}+\alpha R_{ph}^{(1)},
        \end{equation}
        where $R_{ph}^{(0)}=\frac{r_0}{2}\left(1-\sqrt{1-12\mu}\right)$ is a radius of photon sphere in the Vaidya spacetime. 
        The first condition \eqref{eq:condition_potential} being applied to the effective potential \eqref{eq:charged_potential} gives
        \begin{equation} \label{eq:charged_first}
        b^2F(R)\left(1+\alpha G(R)\right) =1,
        \end{equation}
        where
        \begin{eqnarray} \label{eq:def}
        F(R)&\equiv & \frac{1-\frac{2\mu r_0}{R}-2\frac{R}{r_0}}{R^2},\nonumber \\
        G(R)&\equiv &\frac{\nu^2 r_0^2}{R^2\left(1-\frac{2\mu r_0}{R}-2\frac{R}{r_0}\right)}.
        \end{eqnarray}
        Applying the second condition in \eqref{eq:condition_potential} to \eqref{eq:charged_first}, one obtains
        \begin{equation} \label{eq:charged_second}
        F'(R)\left(1+\alpha G(R)\right)+\alpha F(R)G'(R) =0,
        \end{equation}
        where a prime denotes the derivative with respect to the $R$ coordinate. From \eqref{eq:charged_second}, one can easily find that $F'(R_{ph}^{(0)})=0$. Substituting \eqref{eq:charged_radius} into \eqref{eq:charged_second} and expanding in the small parameter $\alpha$, one finds
        \begin{equation} \label{eq:sol}
        F''(R_{ph}^{(0)})\alpha R_{ph}^{(1)}+\alpha G'(R_{ph}^{(0)})F(R_{ph}^{(0)})=0,
        \end{equation}
        which gives
        \begin{equation}
        R_{ph}^{(1)}=-\frac{G'(R_{ph}^{(0)})F(R_{ph}^{(0)})}{F''(R_{ph}^{(0)})}.
        \end{equation}
        Now, by using the definitions \eqref{eq:def} in \eqref{eq:sol}, we can evaluate the sign of $R_{ph}^{(1)}$ to find out if the charge decreases or increases the radius of a photon sphere. For this purpose, we find $F(R_{ph}^{(0)})\,, F''(R_{ph}^{(0)})\,, G'(R_{ph}^{(0)})$ as
        \begin{eqnarray}
        &&F\left(R_{ph}^{(0)}\right)=1-\frac{2\mu r_0}{R_{ph}^{(0)}}-2\frac{R_{ph}^{(0)}}{r_0},\nonumber \\
        &&F''\left(R_{ph}^{(0)}\right)=-\frac{\frac{6\mu r_0}{\left(R_{ph}^{(0)}\right)^2}+\frac{2 R_{ph}^{(0)}}{r_0}}{\left(R_{ph}^{(0)}\right)^3},\nonumber \\
        &&G'\left(R_{ph}^{(0)}\right)F\left(R_{ph}^{(0)}\right)=-\frac{4\nu^2 r_0^2}{\left(R_{ph}^{(0)}\right)^5}.
        \end{eqnarray}
        Substituting these results into \eqref{eq:sol}, one obtains for $R_{ph}^{(1)}$
        \begin{equation} \label{eq:solution1}
         R_{ph}^{(1)}=\frac{4 \nu^2 r_0}{1-12\mu -\sqrt{1-12\mu}},
        \end{equation}
        which is negative for the values of $\mu$ in the interval $\mu \in (0\,, \frac{1}{12})$. Regarding two special values of $\mu$, i.e. $\mu=\frac{1}{12}$ and $\mu=0$, one notes that the Vaidya spacetime admits a homothetic Killing vector if $\mu< \frac{1}{16}$, see the equation (22) in the paper \cite{bib:nel_vaidya} and the discussion thereafter, (also \eqref{eq:vaidya_f2} in section B above). Hence, this condition restricts the possible $\mu$ values to $\mu \in (0, \frac{1}{16})$. Regarding the form of $R^{(0)}_{ph}$ given after (\ref{eq:charged_radius}), we know $R_{ph}^{(0)}\rightarrow 0$ as $\mu \rightarrow 0$. We can show also that $R_{ph}^{(1)}\rightarrow 0$ as $\mu \rightarrow 0$. Here one notes that $ \alpha \nu^2\leq \mu^2$, and hence the limit $\lim\limits_{\mu \rightarrow 0} \alpha R_{ph}^{(1)}$ with  \eqref{eq:solution1} (considering the upper bound of $\alpha \nu^2$ as $\mu^2$)  gives
\begin{equation}
\lim\limits_{\mu \to 0}\alpha R_{ph}^{(1)}=0.
\end{equation}
Moreover, we should also show that $\alpha R_{ph}^{(1)}$ is less than $R_{ph}^{(0)}$, i.e. $R_{ph}$ remains always positive. For this purpose, we let $1-12\mu \equiv y^2$ and consider the upper bound of $\alpha \nu^2$ as $\mu^2$, then \eqref{eq:charged_radius} becomes
\begin{eqnarray}
R_{ph}&=&\frac{r_0}{2}\left(1-y+\frac{(y^2-1)^2}{18y(y-1)}\right) \nonumber\\
&=&\frac{r_0 (y-1)}{36y}\left(-18y+(1+y)^2\right).\end{eqnarray}
  If $0<\mu< \frac{1}{16}$ then $\frac{1}{2}< y<1$ and hence $R_{ph}$ is always positive in this range. $R_{ph}$ vanishes for $y=1$, i.e. $\mu=0$ which is expected because the mass vanishes and there will be no photon sphere.

        {\bf Remark 6:} Here one observes that, like in the Reissner-Nordstrom solution, the electric charge decreases the radius of a photon sphere. Also, similar to the Vaidya solution, the radius of a photon sphere doesn't depend upon the time $t$. 
        
        The critical impact parameter $b_{cr}=\frac{L}{E}$ is then given by
        \begin{equation} \label{eq:charged_impact}
        b=\frac{\left(R_{ph}\right)^2}{R_{ph}^2-2\mu r_0 R_{ph}+\alpha \nu^2 r_0^2-2\frac{R_{ph}^3}{r_0}}.
        \end{equation}
        The angular size of a shadow which can be seen by an observer at $R_o$ is
        \begin{equation} \label{eq:charged_angularsize}
        \sin^2 \omega_{sh}=\frac{b_{cr}^2\left(R_o^2-2\mu r_0R_o+\alpha \nu^2 r_0^2-2\frac{R_o^3}{r_0}\right)}{R_o^4}.
        \end{equation}
        In the old coordinates $(r,v, \theta, \varphi)$ the radius of the photon sphere is given by
        \begin{equation}
        r_{ph}(v)\approx r_{ph}^{V}(v)+\frac{4\alpha \nu^2v}{1-12\mu -\sqrt{1-12\mu}}.
        \end{equation}
        Here we observe the following points. \\
        {\bf Remark 7:} When $R_{o}=R_{ph}$, we have $\omega_{sh}=\frac{\pi}{2}$ and the shadow covers half of the sky. \\
        {\bf Remark 8:} When $R_{ph}<R_o<R_c$, the angular size of a black hole is in the interval $0\leq \omega_{sh}\leq \frac{\pi}{2}$. When the observer position is near the cosmological horizon the  angular size tends to zero, i.e.
        \begin{equation}
        R_o\rightarrow R_c,~~\omega_{sh} \rightarrow 0,
        \end{equation}
        which means a bright sky for the observer.
        \\
        {\bf Remark 9:} When $R_+<R_o<R_{ph}$, we have $\frac{\pi}{2}\leq \omega_{sh}\leq \pi$. When the observer's position is near the outer Killing horizon the angular size tends to $\pi$ 
        \begin{equation}
        R_o\rightarrow R_+,~~ \omega_{sh} \rightarrow \pi,
        \end{equation}
        which means a dark sky for the observer.
        
        Here it is also worth mentioning that the dyonic Vaidya black hole solution \cite{bib:dionic} differs from the Bonnor-Vaidya spacetime with the term $Q_m^2/r^2$ as
\begin{equation} \label{eq:dionic}
ds^2=-\left(1-\frac{2M(v)}{r}+\frac{Q^2_e+Q^2_m}{r^2}\right) dv^2+2dvdr+r^2d\Omega^2
\end{equation}
where $Q_e$ and $Q_m$ are the electric and magnetic charges of the black hole, respectively. As we already mentioned, the homothetic Killing vector exists only for the linear choice of the functions $M(v)$ and $Q_e(v)$. Hence, the magnetic charge affects a photon sphere in the same way as the electric charge does by decreasing it. 

        \section{Naked singularity case}
        
        As one can see from the previous discussion, the shadow doesn't depend upon the conformal factor $e^{\frac{2t}{r_0}}$. Thus, if one considers the Bonnor-Vaidya spacetime in conformally static coordinates, to obtain the shadow, one needs to consider only the static spacetime $d\tilde{s}^2$ \eqref{eq:bvmetric}. This spacetime is a generalization of the Reissner-Nordstrom black hole, and all discussion below will be restricted to this spacetime with the only exception that the notion ''event horizon'' will be replaced by ''conformal Killing horizon''. In this section, we will implicitly show that a naked singularity in Bonnor-Vaidya spacetime may cast a shadow.
        As we pointed out earlier, the radius of a photon sphere in the case of the Reissner-Nordstrom is
        \begin{equation}
        r_{ph}=\frac{3M}{2}\left(1\pm \sqrt{1-\frac{8Q^2}{9M^2}}\right) \,.
        \end{equation}
        Thus, when the ratio $\frac{Q^2}{M^2}$ is in the range
        \begin{equation}
        1< \frac{Q^2}{M^2}\leq \frac{9}{8} \,,
        \end{equation}
        one has two photon spheres in a naked singularity spacetime. However, these photon spheres don't form a shadow.
        We shall prove that in the dynamical case, there is a shadow while in the static Reissner-Nordstrom case, the shadow is absent.
        The apparent horizon location in Bonnor-Vaidya spacetime is given by
        \begin{equation}
    r_{ah}^{\pm}=M(v)\pm \sqrt{M^2(v)-Q^2(v)}.
        \end{equation}
        If $|Q(v)|>M(v)$, the apparent horizon disappears and as a consequence, the spacetime contains a naked singularity. For the linear mass and charge functions, this condition reads as $\mu^2 <\alpha \nu^2$. By denoting
        \begin{eqnarray}
        Q^2&\equiv & \alpha \nu^2 r_0^2,\nonumber\\
        M&\equiv &\mu r_0,
        \end{eqnarray} 
        this condition becomes $M<|Q|$.
        First of all, we shall prove that the generalized Reissner-Nordstrom solution can contain three conformal Killing horizons even if the mass parameter $M$ is less than charge $Q$. Introducing a small parameter $\delta$ by
        \begin{equation}
        Q^2=M^2+\delta,
        \end{equation}
        the generalized Reissner-Nordstrom spacetime reads
        \begin{eqnarray}
        ds^2&=&-f(R)dt^2+2dtdR+R^2d\Omega^2, \nonumber\\
        f(R)&=&\left(1-\frac{M}{R}\right)^2+\frac{\delta}{R^2}-\frac{2R}{r_0}.
        \end{eqnarray}
        One notices that when $\delta=0$, i.e. $M=|Q|$, two conformal Killing horizons do not coincide. It can be seen, for example, from the fact that $f(M)<0$ and $f(2M)>0$ if $r_0>16M$. 
        Letting $\delta>0$  and $r_0=20M$, then $f(M)<0$ if $\delta<\frac{M^2}{10}$. Moreover, since $f(2M)>0$ there will be three horizons: two conformal Killing horizons $R_1$-inner and $R_2$-outer, and one cosmological horizon $R_c$,  $R_c>R_2$. The physical observer is located in $R\in (R_2, R_c)$.
        Thus, we have shown that  $Q^2>M^2$ may lead to a singularity covered by two horizons in the case of the generalized Reissner-Nordstrom spacetime.
        By using the second condition for the photon sphere existence i.e. $V_{eff}'\left(R_{ph}\right)=0$ one obtains
        \begin{equation}
        G(R)\equiv \frac{3M}{R}-\frac{2M^2}{R^2}-\frac{2\delta}{R^2}+\frac{R}{r_0}-1 =0.
        \end{equation}
        Here, we need to prove that this equation has 3 real positive roots. In this case, two photon spheres will be near the conformal Killing horizons $R_\mp$, and the third one near the cosmological horizon. Then we should show that one of the photon spheres is inside the conformal Killing horizon and another one is outside. Hence,  we prove that there will be a shadow. 
        Considering the particular case $r_0=20M$, we need to show that $G(M)>0$, $G(2M)>0$, and $G(R)$ can be negative at $R>2M$. This means that one of the photon spheres is inside and the other one is located outside the conformal Killing horizon.
        It is easy to show that if $r_0=20M$ and $\delta<\frac{M^2}{20}$ then $G(M)>0$ and $G(2M)>0$. However, $G(3M)<0$, so, we can conclude that there is a photon sphere located in the region $2M<R<3M$. Hence we have the following remark.\\
        {\bf Remark 10:}
        The naked singularity in Bonnor-Vaidya spacetime may cast a shadow because in conformally static coordinates it behaves like the generalized Reissner-Nordstrom spacetime. This spacetime, in the case $Q^2>M^2$ may contain three conformal Killing horizons. One of the photon spheres is located in the region where the conformal Killing vector $\frac{\partial}{\partial t}$ is timelike and the other one is in the region where this vector is spacelike. Hence, the physical observer located between the horizons $R_2$ and $R_C$ will observe a shadow.
        
        \section{Conclusion}
        
        In this paper, we have considered the influence of the charge on a dynamical photon sphere. As a model, we have used the Bonnor-Vaidya solution \eqref{eq:chargedvaidya}. This solution as a dynamical spacetime with general mass $M(v)$ and charge $Q(v)$ functions admits only one conserved quantity, the angular momentum-per-mass $L$. Thus,  one has to calculate the shadow properties numerically. However, for the linear choice of mass and charge functions, the Bonnor-Vaidya spacetime admits the homothetic Killing vector which allows the introduction of an extra conserved quantity along the null trajectories. In this case, one can reduce the second-order geodesic equations to the first-order. 
        
        We have used the coordinate transformation to transform Bonnor-Vaidya spacetime to conformally-static coordinates. In this representation, the metric leading to the line element $d\tilde{s}^2$ represents a generalized Reissner-Nordstrom solution. For this spacetime, the radius of a photon sphere has been calculated and compared with the Vaidya solution. We have found out that a charge always decreases the radius of a photon sphere. We have also obtained the angular size of a shadow which can be seen by an observer in the region where the homothetic Killing vector $\frac{\partial}{\partial t}$ is timelike. When the position of an observer is near the horizon $R_C$, which can be interpreted as a cosmological horizon, (s)he sees a bright sky. 
        
        It is known that the naked singularity in the Reissner-Nordstrom case can not cast a shadow because it admits two photon spheres. However, if one considers the dynamical Bonnor-Vaidya spacetime, a naked singularity in this spacetime may cast a shadow. As a consequence, considering a gravitational collapse model leading to a naked singularity, the observer will not be able to distinguish a naked singularity from a black hole through its shadow. This brings the following questions which will be our future research topics.
        \begin{itemize}
        \item Can other dynamical naked singularity models cast a shadow?
        \item Is the shadow of a naked singularity in dynamical spacetime the manifestation of the conformal symmetry only?
        \end{itemize}
        
{\bf Acknowledgments:} V. Vertogradov thanks the Basis Foundation (grant number 23-1-3-33-1) for the financial support.

        
        \section{Appendix: The Homothetic Killing vector}
        
        In a static spherically symmetric spacetime, one can reduce the second-order geodesic equations to the first-order using the conserved quantities,  the angular momentum-per-mass and energy-per-mass. However, in general dynamical spherically symmetric spacetimes there is only one conserved quantity, the angular momentum-per-mass. If the spacetime admits conformal or homothetic Killing vectors, one can introduce an additional constant of motion along the null geodesics. Here one notices that if the spacetime admits a conformal Killing vector, then there exists an additional conserved quantity only along the null geodesics. However, if the spacetime admits the homothetic Killing vector, then there will be an additional conserved quantity not only along the null geodesics but also along the timelike ones.
         Considering the following equation
        \begin{equation}
K_{i;j}+K_{j;i}=C(x^l)g_{ij},~~i,j,l=0,...,3
        \end{equation}
            \begin{itemize}
        \item $K^i$ is called as the Killing vector if $C(x^l)\equiv 0$;
        \item $K^i$ is called as the homothetic Killing vector if $C(x^l)\equiv const.\neq 0$. Homothety means that the line element (with a linear mass function) scales upon a scaling of the coordinates by an overall factor, i.e, \[(v, r)\to (\alpha v, \alpha r)\implies ds^2\to\alpha^2 ds^2\] for any real $\alpha$. As a consequence of this symmetry, if $\left(v(\lambda), r(\lambda)\right)$ is a solution to the geodesic equations in $\lambda$ affine parameter, then  $\left(\alpha v(\lambda), \alpha r(\lambda)\right)$ will be also a solution.
        \item $K^i$ is called as the conformal Killing vector if $C$ is an arbitrary function of $x^l$.
        \end{itemize}
           Now, let's prove that conformal or homothetic Killing vectors can lead to an additional conserved quantity. Let $K^i$ be the conformal or homothetic Killing vector. Then the quantity
        \begin{equation}
        Q\equiv K_i u^i,~~i=0,...,3
        \end{equation}
        is conserved along the null geodesics since
        \begin{eqnarray}
        D_\lambda (Q)&=&D_\lambda (K_i u^i)=\frac{1}{2}\left(K_{i;j}+K_{j;i}\right)u^i u^j\nonumber \\
&=&\frac{1}{2}C(x^l)g_{ij}u^i u^j=0.
        \end{eqnarray}

        \end{document}